\documentclass[a4paper,11pt]{article}
\pdfoutput=1 % if your are submitting a pdflatex (i.e. if you have
\usepackage{jheppub} % for details on the use of the package, please
\usepackage[T1]{fontenc} % if needed
\usepackage{multirow}

\newcommand\aNLO{{\sc\small MadGraph5\_aMC@NLO}}
\newcommand{\ME}[1]{\langle | \mathcal{M}_{#1}|^{2} \rangle}
\newcommand{\NJet}{\textsc{NJet}}
\newcommand{\FastJet}{\texttt{FastJet}}
\newcommand{\Keras}{\texttt{Keras}}
\newcommand{\TensorFlow}{\texttt{TensorFlow}}
\newcommand{\EarlyStopping}{\texttt{EarlyStopping}}
\newcommand{\ReduceLROnPlateau}{\texttt{ReduceLROnPlateau}}
\DeclareMathOperator{\arcsinh}{arcsinh}

\title{
    \vspace{-4cm}\hfill {\small IPPP/21/11 \\ \vspace{-0.3cm} \hfill }\vspace{3cm}\\ 
    A factorisation-aware Matrix element emulator}

\author[a]{D. Ma\^{\i}tre,}
\author[a, b, 1]{H. Truong, \note{Corresponding author.}}

\affiliation[a]{Institute for Particle Physics Phenomenology, Durham University, Durham DH1 3LE, UK}
\affiliation[b]{Institute for Data Science, Durham University, Durham DH1 3LE, UK}

\emailAdd{daniel.maitre@durham.ac.uk}
\emailAdd{henry.truong@durham.ac.uk}

\abstract{
    In this article we present a neural network based model to emulate matrix elements. This model improves on 
    existing methods by taking advantage of the known factorisation properties of matrix elements. In doing so 
    we can control the behaviour of simulated matrix elements when extrapolating into more singular regions than 
    the ones used for training the neural network. We apply our model to the case of leading-order jet production 
    in $e^+e^-$ collisions with up to five jets. Our results show that this model can reproduce the matrix elements
    with errors below the one-percent level on the phase-space covered during fitting and testing, and a robust
    extrapolation to the parts of the phase-space where the matrix elements are more singular than seen at the
    fitting stage.
}

\begin{document} 
\maketitle
\flushbottom

\section{Introduction}
\label{sec:intro}

Neural networks (NN) first appeared in the field of particle physics as a tool for event discrimination in the analysis of collider data. 
Since then neural networks and other machine learning techniques have proved useful in many other areas of the field. 
On the theory prediction side they have been used 
to improve the efficiency of Monte Carlo sampling \cite{Bendavid:2017zhk,Klimek:2018mza,Bothmann:2020ywa,Verheyen:2020bjw,Chen_2021}, 
to accelerate the simulation of radiation within a jet~\cite{Carrazza:2019cnt,Bothmann:2018trh,Dohi:2020eda}, 
to streamline the processes of generation and unweighting of simulated event samples and 
\cite{Gao:2020zvv,Otten:2019hhl,Hashemi:2019fkn,DiSipio:2019imz,Butter:2019cae,Bishara:2019iwh,Backes:2020vka,Butter:2020qhk,Alanazi:2020klf,Nachman:2020fff}.
Closer to the experimental measurements 
they have also been used to emulate detector simulation \cite{Paganini:2017dwg,SHiP:2019gcl,Derkach:2019qfk,Alanazi:2020jod},
 they can be used to perform unfolding \cite{Andreassen:2019cjw} or correcting for detector effects \cite{Bellagente:2019uyp},
and perform pileup subtraction \cite{Andreassen:2019cjw}. 

The capacity of neural networks to approximate intricate functions have already been 
used to provide fast calculations of production cross-sections~\cite{otten2019deepxs,Buckley:2020bxg}.
In this article we investigate whether NNs can approximate production cross-sections more differentially by replacing computationally expensive 
matrix element calculations. 
The challenge with this endeavour is that matrix elements are plagued with numerous divergences that arise from infrared divergences.
In previous works \cite{Badger:2020uow,Aylett-Bullock:2021hmo} a combination of individual neural networks were used 
to approximate matrix elements. In order to deal with the complex structure of the matrix elements 
the authors of these studies divided the phase-space into sectors according to the infrared singularities and trained networks on these sectors,
thereby limiting the complexity of the fit by isolating a single divergence per sector. All sectors were then combined to make the final prediction.
While the authors of the initial study for electron-positron annihilation showed good agreement between the total cross-section and histogrammed distributions both at LO and NLO,
we note that the accuracy of the interpolation at the level of individual points was a lot worse than when averaged in the 
histograms. In addition, the performance of the extrapolation outside of its training phase-space 
(i.e. more singular configurations than those considered to fit the model) is problematic.   

In this article we present a different approach to the emulation of matrix elements that incorporates the factorisation 
properties of the matrix elements in the interpolation model and therefore is able to safely 
extrapolate the matrix element in regions more singular than that covered by the data used for the training of the emulator. We find
that our interpolator also displays a much improved pointwise accuracy.

The article is organised as follows. 
Section~\ref{sec:fitting} introduces our factorisation-aware 
deep neural network model, Section~\ref{sec:results} showcases our results with a comparison with the 
model from Ref.~\cite{Badger:2020uow} in Section~\ref{sec:n3jet_comparison}, and an analysis of the performance of our model in Section~\ref{sec:mainresult}. 
To show that the neural network is both interpolating and extrapolating well we show its 
behaviour on random trajectories in phase-space in Section~\ref{sec:trajectories}. 
Our conclusions are presented in Section~\ref{sec:conclusion} 
and some discussion of specific details are collected in the appendices so as not to distract from the main discussion.

We provide code to reproduce the methodology detailed in this article \cite{fame_repo}.

\section{Fitting framework}\label{sec:fitting}
For this work we consider the $e^+e^-\rightarrow Z^{*} / \gamma \rightarrow q\bar q +n g$ matrix elements for $n$ up to and including 3, 
which corresponds to events with up to 5 jets. We denote the number of jets in the final state as $n_{j}$.

We formulate the problem of emulating matrix elements as a supervised regression task with a set of phase-space points' kinematic information as input and the 
values of the matrix element for each of these phase-space points as the targets. Section~\ref{sec:data} describes how these matrix elements were obtained.

A neural network can be seen as a function $f(\vec{x}; \theta) = \vec{y}$, where $f: \mathbb{R}^{d} \rightarrow \mathbb{R}$ maps a 
$d$-dimensional vector $\vec{x}$ of inputs onto the vector of outputs $\vec{y}$, and where $\theta$ are the parameters of the neural network which we aim to optimise 
such that the outputs $\vec{y}$ of the neural network match the target as well as possible. 
The simplest implementation of an emulator would be to take the input as the kinematic information of the phase-space point and the output to be the full matrix element. 
Our approach modifies both the input of the NN and its target, as described in the following sections.

\subsection{Infrared divergences and dipole factorisation formula}
It is well known that in soft and collinear limits the matrix element in $n+1$-body phase-space factorises into a singular factor and a reduced matrix element in $n$-body phase-space \cite{ALTARELLI1977298,BASSETTO1983201}.
This factorisation was used by Catani and Seymour \cite{Catani:1996vz} to construct subtraction terms for the real radiation part of a NLO calculation.
They introduced a factorisation formula with universal dipoles that smoothly interpolates 
between the soft and collinear limits to capture the singular structure in these regions of phase-space.
The dipole factorisation formula can be written schematically as
\begin{eqnarray}\label{eq:factorisation}
    |\mathcal{M}_{n+1}|^{2} \rightarrow |\mathcal{M}_{n}|^{2} \otimes \mathbf{V}_{ij,k} \, ,
\end{eqnarray}
where $\mathbf{V}_{ij,k}$ is a process independent, singular factor. It depends on the momenta and quantum numbers (colour and spin) of partons $i,j,k$, where $i$ is the emitter parton, 
$j$ is the emitted parton, and $k$ is the spectator parton.
For singly unresolved limits, this factorisation isolates all the divergent behaviour in $\mathbf{V}_{ij,k}$ and the factor $|\mathcal{M}_{n}|^{2}$ is free of divergences, which makes 
it more amenable to emulation through a neural network.
The dipole factorisation formula forms the basis of our fitting ansatz which we present in detail in Section~\ref{sec:ansatz}.

\subsection{Fitting coefficients of Catani-Seymour dipoles}\label{sec:ansatz}
\begin{figure}
    \includegraphics[width=\textwidth]{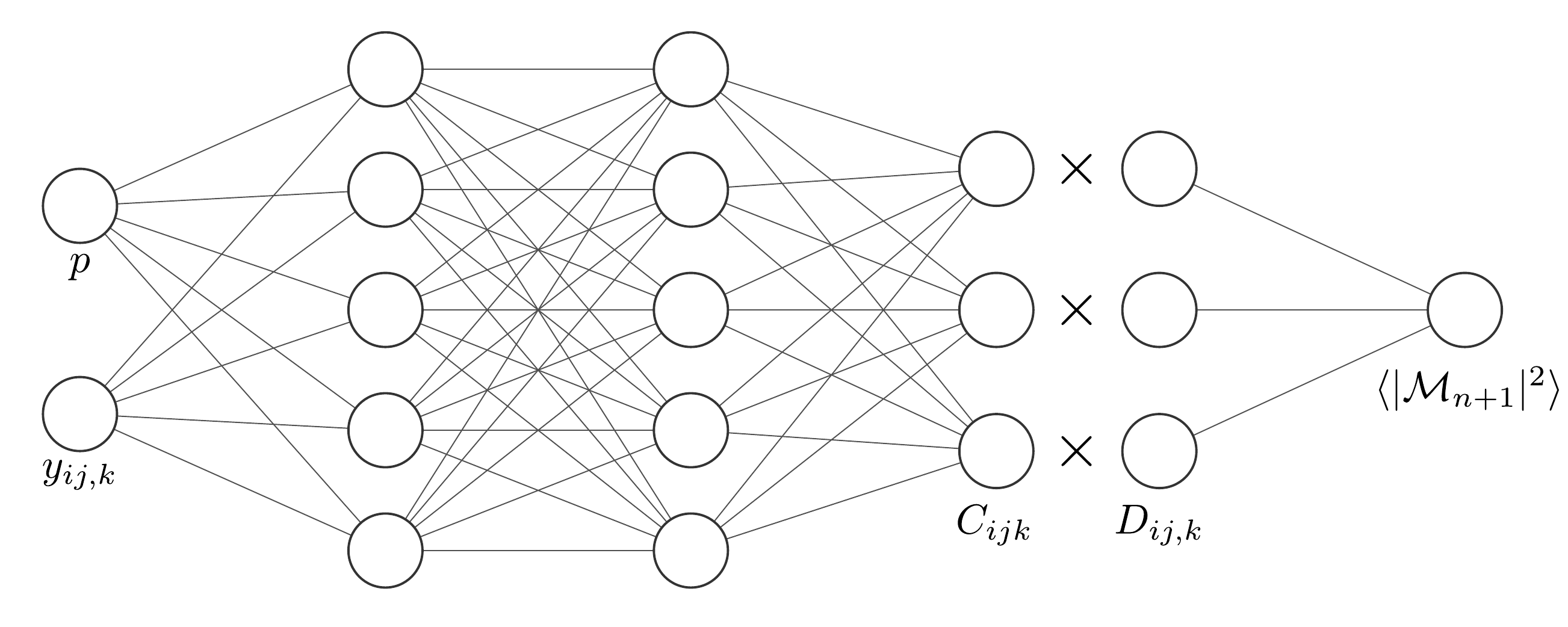}
    \caption{Schematic diagram of our neural network architecture.
    We have a densely-connected neural network with inputs phase-space points, $p$, and recoil factors, $y_{ij,k}$, propagated through hidden layers to the output layer which outputs $C_{ijk}$.
    These coefficients are combined with their corresponding spin-averaged dipoles as in (\ref{eq:ansatz}) to produce an approximation of the matrix element.
    Diagram of neural network generated with the aid of \cite{LeNail2019}.}
    \label{fig:nn_architecture}
\end{figure}

Instead of using a neural network to fit the matrix element directly, we use the dipole factorisation formula to build an ansatz of the colour and helicity summed $n+1$-body matrix element,
\begin{eqnarray}\label{eq:ansatz}
    \ME{n+1} = \sum_{\{ijk\}} C_{ijk} D_{ij,k} \, ,
\end{eqnarray}
where $D_{ij,k} = \langle V_{ij,k} \rangle / s_{ij}$ are the spin-averaged Catani-Seymour dipoles divided by the corresponding Mandelstam invariant and $C_{ijk}$ are the coefficients we train the neural network to fit.
$C_{ijk}$ can be interpreted as the reduced matrix element in $n$-body phase-space. 
Since the input for the $C_{ijk}$ function is the full $n+1$ phase-phase information, the neural network will also model the phase-space mappings usually introduced in the factorisation formula.
A schematic diagram illustrating our ansatz is given in Figure~\ref{fig:nn_architecture}.
The sum over ${\{ijk\}}$ denotes the sum over relevant permutations of the external outgoing legs.
More detail on this is given in Section~\ref{sec:architecture}.
The representation (\ref{eq:ansatz}) is not unique but 
through appropriate training, the neural network takes advantage of the right ingredients to model 
the divergent soft and collinear behaviour of the matrix elements.

This form of the ansatz allows the neural network to avoid fitting a rapidly varying function over the phase-space, leaving the Catani-Seymour dipoles to reproduce the correct singular behaviour, meaning a single neural network can interpolate a now relatively smooth function over the phase-space.

\subsection{Data generation}\label{sec:data}
For all multiplicities, phase-space is sampled uniformly using the RAMBO algorithm \cite{Kleiss:1985gy} with a centre-of-mass energy $\sqrt{s_{\mathrm{com}}} = 1000$ GeV.
Phase-space points are subsequently clustered using {\FastJet} \cite{Cacciari:2011ma,noel_dawe_2021_4446849} with the $e^{+}e^{-}$ $k_{t}$ algorithm \cite{Catani:1991hj}.
Global phase-space cuts are applied according to the criterion $y_{\mathrm{cut}} \leq y_{ij}$ where $y_{ij}$ are the Mandelstam invariants normalised by $s_{\mathrm{com}}$.
Jets are clustered exclusively where $d_{\mathrm{cut}}$ was supplied to {\FastJet}. We took $d_{\mathrm{cut}} = \mathrm{max}(2 \times y_{\mathrm{cut}}, 0.01 \times s_{\mathrm{com}})$.
We explore three different values of the global phase-space cut parameter, $y_{\mathrm{cut}} = [0.01, 0.001, 0.0001]$, to demonstrate the ability of the factorisation-aware neural network to effectively interpolate in more and more singular regions of phase-space.

The generated phase-space points are fed to the {\NJet} package \cite{Badger:2012pg} to calculate colour and helicity summed tree-level matrix elements. 
All external legs have been considered to be massless. 
The strong coupling constant has been set to $\alpha_{s} = 0.118$, and the electromagnetic coupling constant has been set to $\alpha_{e} = 1.0 / 132.5070$.
The mass of the $Z$-boson is taken to be $m_{Z} = 91.188$ GeV.
These parameters and those not listed, are consistent with those in the Standard Model mode of \aNLO \cite{Alwall_2014}.

The phase-space points generated form the basis of our inputs to the neural network with the matrix elements as our fitting targets.
As with most machine learning applications, we need to demonstrate that our neural network emulator has managed to generalise outside of the training dataset.
We do this by firstly testing on an independent testing dataset that is never exposed to the network during training, and secondly by predicting on random trajectories in phase-space.
Generation of phase-space trajectories is described in Appendix~\ref{appendix:trajectory_generation}.
We believe that accurate predictions on random phase-space trajectories demonstrates the ability of the neural network to extrapolate to never before seen data 
that is of a different nature to both the training and testing datasets.

\subsection{Neural network emulator}\label{sec:network}
We construct our emulator using a densely-connected neural network built using the {\Keras} API \cite{chollet2015keras} with the {\TensorFlow} back-end \cite{tensorflow2015-whitepaper} with GPU support. A simple model architecture such as a densely-connected neural network allows for quicker training and inference compared to more complicated setups.
\subsubsection{Inputs and outputs}\label{sec:input_outputs}
\paragraph*{Inputs to neural network}
As mentioned in Section~\ref{sec:data}, phase-space points form the basis of the inputs to our neural network.
We input the 4-momenta of each outgoing parton as an $(n_{j} \times 4)$ array with each component of its 4-momenta standardised to zero mean and unit variance across the training dataset.
Although it would be feasible to omit the energy component of the momenta or leave one outgoing parton out of the training due to our datasets being generated with a fixed centre-of-mass energy, 
we find that keeping all momenta information to improve the network's performance.
With the acceleration of training of neural networks on GPUs, the slowdown in keeping all outgoing momenta components is negligible.

Along with the 4-momenta, we also include the recoil factors
\begin{eqnarray}\label{eq:recoil_factor}
    y_{ij,k} = \dfrac{p_{i}p_{j}}{p_{i}p_{j} + p_{j}p_{k} + p_{i}p_{k}} \, ,
\end{eqnarray}
as input for relevant permutations of $\{ijk\}$.
We take the natural logarithm of $y_{ij,k}$ before standardising them to a zero mean and unit variance across the training dataset.
We find that the addition of these recoil factors significantly improves performance of our neural network emulator during training and testing.
This can be attributed to the fact that the coefficients $C_{ijk}$ rely heavily on the mapped momenta in $n$-body phase-space, where $y_{ij,k}$ is usually required when performing the momentum mapping, see e.g. \cite{Catani:1996vz}.

Following our ansatz (\ref{eq:ansatz}), we must provide spin-averaged Catani-Seymour dipoles to the neural network. These dipoles are computed for all phase-space points before training begins, but they are not passed directly to the neural network as input features. Instead, they are only included in our custom loss function which will be explained in the next subsection.

\paragraph*{Accounting for spin-correlation in $g \rightarrow gg$}
In addition to the Catani-Seymour dipoles, we include other functions to account for the spin-correlations of $g \rightarrow gg$ and $g \rightarrow q\bar q$ splittings which are present in the factorisation formula but averaged out in the spin-averaged dipoles.
This effect becomes relevant when there are two or more gluons in the final state.
We seek to capture this behaviour by introducing a pair of terms of the form 
\begin{equation}
S_{ij} \sin\left(2 \phi_{ij}\right) + C_{ij} \cos\left(2\phi_{ij}\right)
\end{equation}
in the fitting ansatz for each gluon pair. The coefficients $S_{ij}$ and $C_{ij}$ are fitted by the neural network along with the dipole coefficients. The angle $\phi_{ij}$ is the azimuthal angle of the decay particles in the plane perpendicular to the parent particle momentum. The procedure to obtain this angle is described in Appendix~\ref{appendix:phi}.

\paragraph*{Outputs of neural network}
Denoting the raw output of our neural networks as $c_{ijk}$, they are transformed to $C_{ijk}$ according to
\begin{eqnarray}\label{eq:coef_output}
    C_{ijk} = S_{\mathrm{coef}} \times \sinh{(c_{ijk})} \,
\end{eqnarray}
where $S_{\mathrm{coef}} = S_{\mathrm{pred}} / S_{\mathrm{dipole}}$.
$S_{\mathrm{pred}}$ is the prediction scale, taken to be the minimum of the matrix elements in the training set and $S_{\mathrm{dipole}}$ is the dipole scale, taken to be the mean of all dipoles in the training set.
These scaled coefficients are then multiplied with their corresponding Catani-Seymour dipole $D_{ij,k}$, and then summed to produce an estimation of the matrix element.
The targets are the matrix elements corresponding to the phase-space point inputs in the training dataset. We transform the targets according to
\begin{eqnarray}\label{eq:arcsinh}
    y = \arcsinh{\left(\dfrac{\ME{n}}{S_{\mathrm{pred}}}\right)} \, ,
\end{eqnarray}
to reduce the orders of magnitude the matrix elements span.
This performs a similar transformation to taking the natural logarithm except that it remains a valid transformation for negative arguments.
We require this transformation to allow negative values as arguments because the coefficients $C_{ijk}$ will not be restricted to only positive values, meaning that during training there is a possibility for the outputs of the network to go negative.
By reducing the span of the targets, the neural network is able to more effectively pick out patterns across the entire training dataset rather than a smaller region, helping it to generalise.
This technique is employed in other studies, see for example \cite{Coccaro_2020,Bury_2021}.
We would like to stress that we do not expect, and have not observed, negative predictions for the matrix element from the neural network, as this would be unphysical.
$y$ is finally standardised to a zero mean and unit variance.

\paragraph*{Training, validation, and testing datasets}
In Ref. \cite{Badger:2020uow}, the authors used 500k training samples to train their models.
For a fair comparison of our respective methods, we follow their methodology closely by constructing our models to be as close as possible to theirs and generate training and testing data by using code from their project repository \cite{n3jet_repo}\footnote{for our main results we generate all data using our own code}.
Our model architectures will not be identical due to the difference in our methods, details of our model architecture are given in Section~\ref{sec:architecture}.
Results of this comparison are presented in Section~\ref{sec:n3jet_comparison}.

While training on 500k samples gives acceptable performance for the total cross-section, per-point accuracy is lacking.
We find that increasing the size of the training dataset drastically improves the per-point prediction accuracy.
Neural networks have been shown to scale well with large datasets \cite{hestness2017deep} and given that they only need to be trained one time, it is useful to provide neural networks which have been pre-trained with maximum accuracy in mind.
In addition to improving the per-point accuracy, we aim to overcome the problem of extrapolating to more singular regions.
It is well known that neural networks in general do not extrapolate well \cite{xu2021neural,kaplan2020scaling}, but with our factorisation-aware model we show that letting the models learn about the infrared structure of QCD alleviates this problem.

To demonstrate the scaling performance of our model, we present our main results with models trained on more training samples where details of data generation are given in Section~\ref{sec:data}.
For each multiplicity and global phase-space cut we generate a dataset consisting of 60 million phase-space points and their corresponding recoil factors, dipoles, and matrix elements.
We then split this dataset into training, validation and testing datasets in a ratio of 4:1:1, meaning we have 40 million training samples, 10 million validation, and 10 million testing samples.
The validation dataset is used to monitor model performance after each epoch of training, whereas the testing dataset is used as an out-of-sample check of the model's performance after training is complete.

\subsubsection{Architecture}\label{sec:architecture}
We have a fixed base to our neural network architecture that is used for all processes considered in this work, with variations in the number of nodes in the input and output layers due to the change in number of outgoing partons.
Our base neural network consists of one input layer, eight hidden layers consisting of (64, 128, 256, 512, 768, 386, 128, 64) nodes, and one output layer.

The number of nodes in the input and output layers scales with the number of jets in the final state.
The input layer has $(n_{j} \times 4) + n_{\mathrm{rel}}$ nodes, and the output layer has $n_{\mathrm{rel}} + (2 \times n_{\phi})$ nodes, where $n_{\mathrm{rel}}$ is the number of relevant permutations and $n_{\phi}$ denotes the number of $\phi_{ij}$ angles.
Relevant permutations, ${\{ijk\}}$, are the set of permutations for which their corresponding dipole could have the possibility to have a meaningful contribution to the matrix element for a given process.
They are a subset of all the possible permutations for a given multiplicity, $P(n_{j}, 3)$.
Relevant permutations exclude any permutation where a quark or anti-quark are emitted (i.e. $j = q$ or $\bar{q}$), as low energy quarks do not give rise to singularities in our processes of interest.
Furthermore, we can remove degenerate permutations where swapping $i=g$ and $j=g$ has no effect as they have identical Catani-Seymour dipoles, e.g. $D_{34, 1} = D_{43, 1}$, so we only keep $D_{34, 1}$.
The omission of these redundant permutations speeds up training of the neural networks as we have fewer inputs, fewer dot products to compute in the loss function, as well as speeding up inference due to there being fewer dipoles to compute.
For reference, we list the number of input and output nodes for each multiplicity we consider in Table \ref{table:nodes}.

\begin{table}
    \centering
    \begin{tabular}{|c|c|c|}
        \hline
        Number of final state jets & Input nodes & Output nodes \\
        \hline
        3 & $(3 \times 4) + 2  = 14$ & $2 + 2 \times 0  = 2$   \\
        4 & $(4 \times 4) + 10 = 26$ & $10 + 2 \times 1 = 12$ \\
        5 & $(5 \times 4) + 27 = 47$ & $27 + 2 \times 3 = 33$ \\
        \hline
    \end{tabular}
    \caption{List of the number of input and output nodes for every process we consider.}
    \label{table:nodes}
\end{table}

The neural network weights are initialised according to the `Glorot uniform' distribution as described in \cite{Glorot10understandingthe}.
We use the $\tanh$ activation function for all nodes in the hidden layers, and have a linear activation function for nodes in the output layer.
Initial learning rate is set to 0.001 and training mini-batch size is set to 4096.
During training, we reduce the learning rate by a factor of $0.7$ whenever there is no improvement in validation loss for 20 epochs.
We use the {\Keras} callback {\ReduceLROnPlateau} to achieve this.
Model training is terminated after the validation loss does not improve after 40 epochs of training using the {\EarlyStopping} callback in {\Keras}.
We find that reducing the learning rate during training helps the model to converge to more optimal parameter sets.
Since there are periods during training where the validation loss stagnates, reducing the learning rate helps to reach minima which otherwise wouldn't be accessible due to a too large learning rate.
There is a possibility to reduce the learning rate too rapidly causing the model to have suboptimal optimisation, but this is countered by the high patience we set for the {\ReduceLROnPlateau} callback.

These choices of hyperparameters were not results of extensive scanning of parameter space and were chosen heuristically.
Hyperparameters can be tuned more optimally using more sophisticated methods, but they all rely on training a large number of models which is computationally expensive and time-consuming.

\paragraph*{Custom loss function}
To assess the model's performance when training we need to compare the network predictions with the targets. Our metric for the model's regression performance is the mean squared error (MSE)
\begin{eqnarray}\label{eq:mse}
    L_{\mathrm{MSE}} = \dfrac{1}{N} \sum_{i=1}^{N} \left(y_{i} - p(\vec{x}_{i}; \theta)\right)^{2} \, ,
\end{eqnarray}
where $y_{i}$ is the target for the $i$-th sample out of $N$ samples, and $p(\vec{x}_{i}; \theta)$ is the corresponding prediction
obtained by combining the dipole factors and the azimuthal dependency terms multiplied by their NN-learned coefficients. 
In order to correct for the scale difference, we need to apply the transformation (\ref{eq:arcsinh}) on the neural network prediction first, with the same $S_{\mathrm{pred}}$ that was used for scaling the targets.
We choose the mean squared error to measure the network's performance because it is sensitive to outliers in the target distribution.
This is useful because even though we have taken measures to reduce the span of the targets, the distribution still contains a tail towards larger values which correspond to soft and collinear configurations.
These points have large contributions to the cross-section when integrating over phase-space, meaning it is important that we accurately predict these points.
It is also convenient that the mean squared error tends to learn the mean of the target distribution which, in our case, corresponds to the cross-section.

In addition to the MSE, we introduce a regularisation term to penalise non-sparse representations of the matrix element. We know that in soft and collinear limits there will be dominant dipoles that have large contributions to the matrix element and there will be other dipoles with minimal contribution. We try to suppress the coefficients associated with these minimally contributing dipoles with the penalty term
\begin{eqnarray}\label{eq:penalty}
    L_{\mathrm{pen}} = J \, \sum_{i} \dfrac{D_{i}^{-2}}{\sum_{j}D_{j}^{-2}} |C_{i}D_{i}| \,
\end{eqnarray}
where the sum over $i$ replaces the sum over $\{ijk\}$ for brevity, and $\sum_{j} D_{j}^{-2}$ is the sum over all dipoles for a phase-space point, acting as a normalisation factor.
$J$ is a tunable parameter that scales the importance of $L_{\mathrm{pen}}$ versus $L_{\mathrm{MSE}}$.
We found that models perform the best when $L_{\mathrm{pen}} < L_{\mathrm{MSE}}$, so $J$ is tuned accordingly.
It is possible for the product $C_{i}D_{i}$ to be large due to $D_{i}$ alone, so to penalise this we regularise the product rather than just the coefficient, since it is the product that contributes to the matrix element.

The form of $L_{\mathrm{pen}}$ is reminiscent of the usual L1 regularisation which promotes sparse models.
Regularisation is usually included to prevent overfitting by making the model make decisions on the most important features, reducing other features to zero.
In our case we would like the neural network to learn about the universal factorisation property in QCD by making it choose a minimal amount of dipoles to represent the matrix element in singular regions.
In addition to preventing overfitting, (\ref{eq:penalty}) helps the neural network to extrapolate to more soft and collinear regions as it has learnt to choose which dipoles are relevant in specific configurations.
For non-singular configurations, the neural network is free to interpolate as there is not a clear set of dipoles that dominate, meaning $L_{\mathrm{pen}}$ is small.

Combining the regression loss term and the regularisation term, our expression for the total loss is
\begin{eqnarray}\label{eq:loss}
    L = L_{\mathrm{MSE}} + L_{\mathrm{pen}} \, ,
\end{eqnarray}
which is minimised through mini-batch gradient descent \cite{bottou2018optimization} with the Adam optimiser \cite{kingma2017adam} to find optimal parameters $\theta$ for the neural network.

\paragraph*{Ensemble of models}
Due to the random initialisation of weights in the neural network, and the fact that the optimisation procedure is carried out on mini-batches of the full training dataset, every neural network trained will be similar but non-identical, even with identical model architecture.
This is partly because the loss surface is unlikely to be a completely smooth surface with a single global minimum, instead it is likely to contain multiple local minima, meaning it would be optimistic to believe that a single neural network is able to find the most optimal set of parameters.
Given that we don't expect a single network to perform optimally, we train a number of models and aggregate their predictions to create an ensemble of models.
Ensembling models is a well-known technique within machine learning, for a review see \cite{ZHOU2002239}.

Each model in the ensemble is initialised with different weights according to the `Glorot uniform' distribution, and trained on the same but randomly shuffled dataset, resulting in models that have been exposed to different distributions of the training data.
After sufficient training, each model will have a distinct set of parameters that have similar predictive power.
The prediction for the matrix element is then the mean of the outputs of all models in the ensemble\footnote{if not explicitly stated, all references to neural network predictions from henceforth refers to the prediction from the ensemble of networks}.
Taking the mean will give a more accurate and robust prediction as averaging over the different models will reduce variance due to over/underfitting in the training phase.
We choose to have 20 models in our ensembles because we begin to see diminishing returns in the accuracy of per-point predictions after this.
Another advantage of training an ensemble of models is that we have a measure of uncertainty, due to the neural networks, on the model predictions by calculating the standard error of the mean.
That is we take the standard deviation of predictions across the ensemble and divide by the square root of number of models in the ensemble. This would not be possible with just a single model.

By choosing to build an ensemble, there is a performance impact because we have to spend more resources on training, and inference is slower due to having to predict on all models in the ensemble.
However, we believe that having a more robust prediction with a measure of uncertainty outweigh these negatives.
Additionally, each model only needs to be trained once, and slowdown during inference is alleviated with GPUs. 

\section{Results}\label{sec:results}
In this section we present results for our matrix element emulator for $e^{+}e^{-}$ annihilation into up to 5-jets.
We first compare results obtained with our method to the tree-level results of Ref.~\cite{Badger:2020uow}, then proceed to exhibit results from larger NNs
that have been exposed to larger training datasets to demonstrate the full scaling performance of our model.
We then further demonstrate our method's capability to generalise to unseen regions of phase-space by assessing the prediction accuracy on random phase-space trajectories that venture well outside of the phase-space region used for the training.

\subsection{Comparison with previous work}\label{sec:n3jet_comparison}
We compare our method to methods for tree-level matrix elements emulation from Ref.~\cite{Badger:2020uow}.
The authors presented two methods: `single' and `ensemble' models. A `single' models indicates that there is 
one neural network trained across the entire phase-space, and an `ensemble' model indicates that there is a group of neural networks trained together
with weighting functions that focus individual networks on a specific divergent region of phase-space.
In order to conduct a fair comparison, we have made attempts to follow their training methodology closely by 
constructing the neural networks at the centre of our model with a similar structure (e.g. same hidden layer structure, same activation functions, 
same random initial weight distribution), training methodology (same {\EarlyStopping} criteria, same initial learning rate), and have generated training and testing datasets using code 
available on the authors' project repository with the relevant cuts.

\begin{figure}
    \centering
    \includegraphics[width=\textwidth]{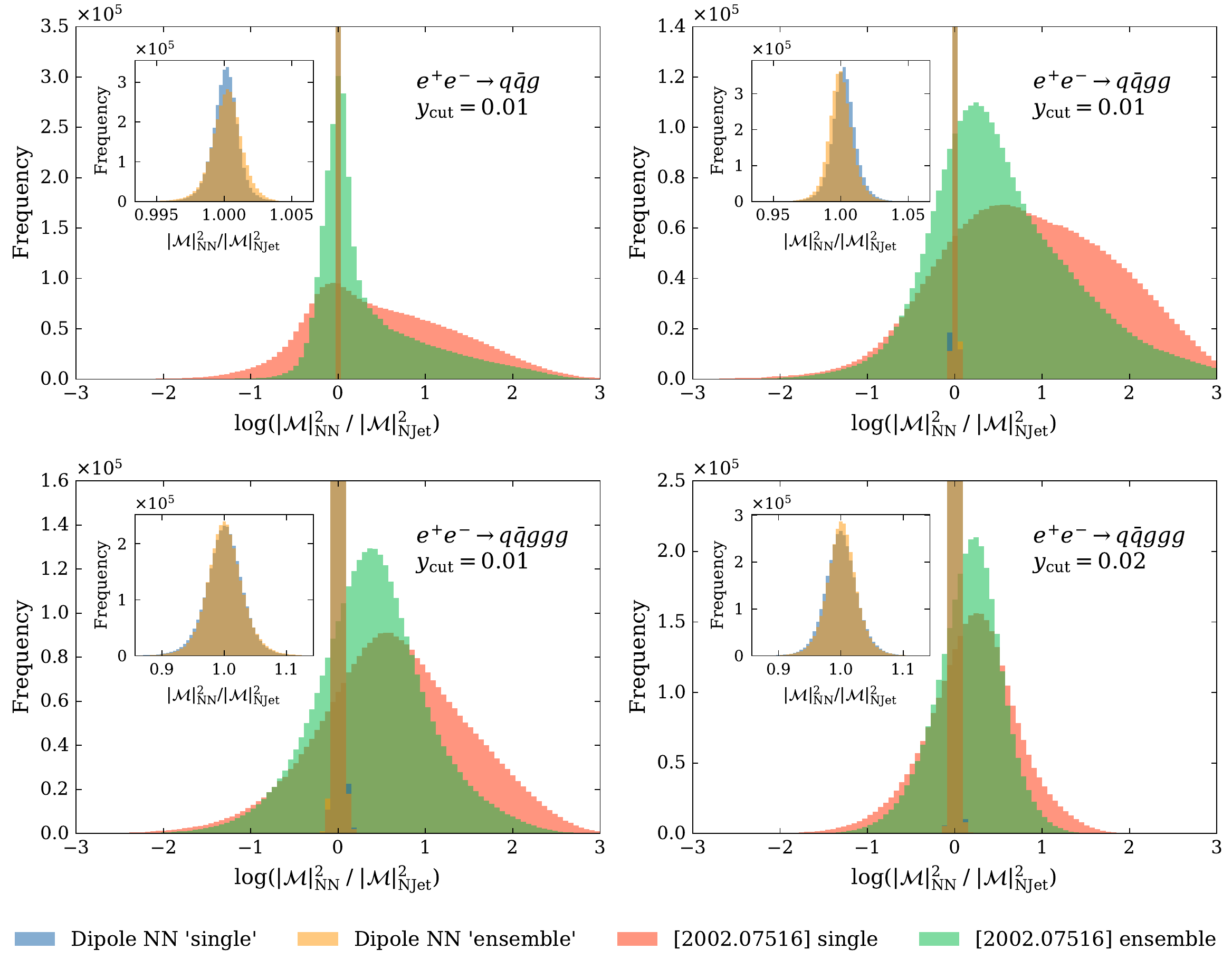}
    \caption{Error distribution compared to Figure 3 in Ref.~\cite{Badger:2020uow}, where data to reproduce the histograms were provided by the authors.
    We plot the log ratio of the matrix element as predicted by the neural network ensemble and the value from {\NJet} on the main axes for comparison. The blue and orange dipole histograms representing our method 
    are cut off at the top on the main axes, but the most important feature is the narrowness of the peak centred around the ideal value.
    The insets show the detailed distribution of our result on a linear scale.}
    \label{fig:comparison}
\end{figure}

We compare the distribution of errors for the matrix elements predictions on 3 million phase-space points.
The distribution of errors is crucial because it informs us of the performance of our emulators at the matrix element level.
It is well known that a neural network optimised with the mean squared error loss function has tendencies to learn the mean of the target distribution \cite{Nachman_2020},
meaning the quality of the cross-section prediction can potentially belie the point-to-point accuracy of emulators.
In Figure \ref{fig:comparison}, we plot the distribution of errors for matrix element predictions where our method is labelled `Dipole NN'.
We compare against the `single' and `ensemble' methods by training and testing using the corresponding datasets.
For more detailed information on the differences please refer to \cite{Badger:2020uow}.
Note that the height of the peak for the dipole histograms are not illustrated in the figure as it would not fit on the current axes but that is not important for this comparison.
We can see that the prediction-to-truth ratio distribution for our method is much narrower and consistently peaked around the ideal accuracy, indicating our model performs better on a per-point basis for all multiplicities.
Even with this reduced NN size we can see that incorporating the known
divergent structure explicitly in the model gives better results, as it
uses the NN representation to learn a function that is more
suitably approximated by a NN.
For example, even though the three jet matrix element has a fairly trivial analytical structure, a standard
fitting approach using a NN typically struggles to reproduce
divergences. In our approach the NN only needs to emulate a non-singular
modulation on top of the main divergent behaviour and is therefore more
suited to the task.

\subsection{Main results}\label{sec:mainresult}
\begin{figure}
    \centering
    \includegraphics[width=\textwidth]{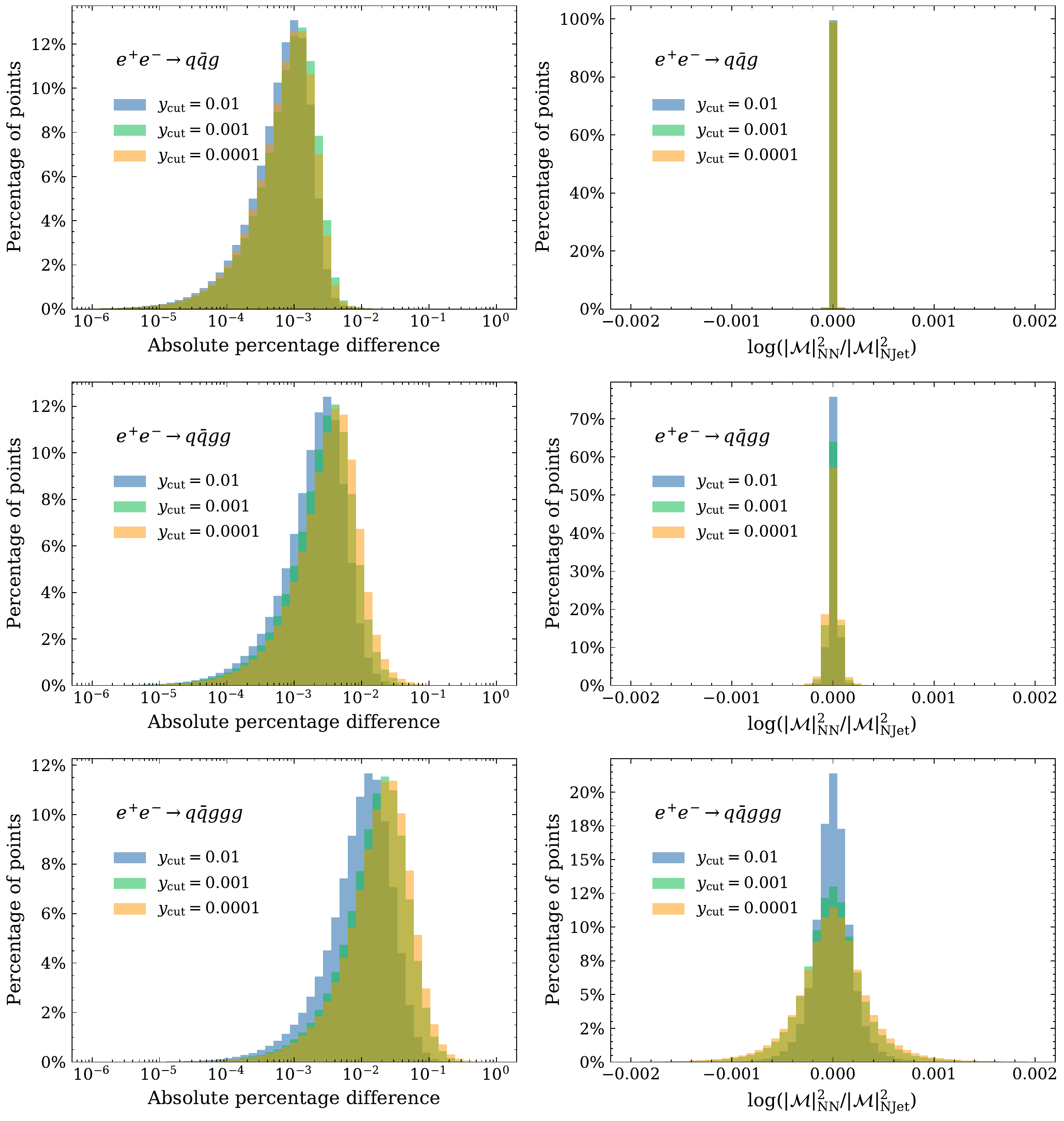}
    \caption{Error distributions for all three multiplicities (rows) and global phase-space cuts.
    Left: absolute percentage difference between NN prediction and {\NJet}.
    Right: ratio of matrix elements from NN and {\NJet}.
    Axis scales have been fixed for each column of subplots for ease of comparison between multiplicities.}
    \label{fig:main_error_distribution}
\end{figure}

Here we present our main results which are obtained using the larger NNs described in Section~\ref{sec:architecture} along with larger training datasets described in Section~\ref{sec:data}.
In Figure~\ref{fig:main_error_distribution}, we show the error distributions on 10 million matrix element predictions for each multiplicity and global phase-space cut.

Predicting on a large number of phase-space points allows us to explore singular regions with higher statistics. 
The ratios (right) clearly highlight the symmetry of the errors with Gaussian-like distributions tightly centred around 0.
We have also included the absolute percentage difference distribution for more easily interpretable errors where the bulk of predictions are below the 0.1\% error level.
With increasing multiplicity, the fitting gets more challenging due to the rise in the number of singular regions in phase-space and the dimensionality of the phase-space, which can be seen in the decrease in accuracy as we increase multiplicity.
Although the errors do increase, practically all matrix element predictions are below the 1\% error level even for the most challenging scenario.
Relaxing the global phase-space cut for the training and testing set is also expected to decrease model performance as allowing more singular regions of phase-space stretches the span of the target distribution making it difficult to fit with a single neural network.
Our method manages to retain good performance while global phase-space cuts have been relaxed by a factor of 100 with only a small decrease in accuracy as illustrated in the left column of Figure~\ref{fig:main_error_distribution}.
This is because most of the span is accounted for by the dipole factors, while the coefficients themselves vary less.
In the 3-jet case there is negligible difference between the different phase-space cuts while the 4 and 5 jet cases see less than a factor of 10 difference in the peaks of the absolute percentage difference distributions going from $y_{\mathrm{cut}} = 0.01$ to $y_{\mathrm{cut}} = 0.0001$.

\begin{figure}
    \centering
    \includegraphics[width=\textwidth]{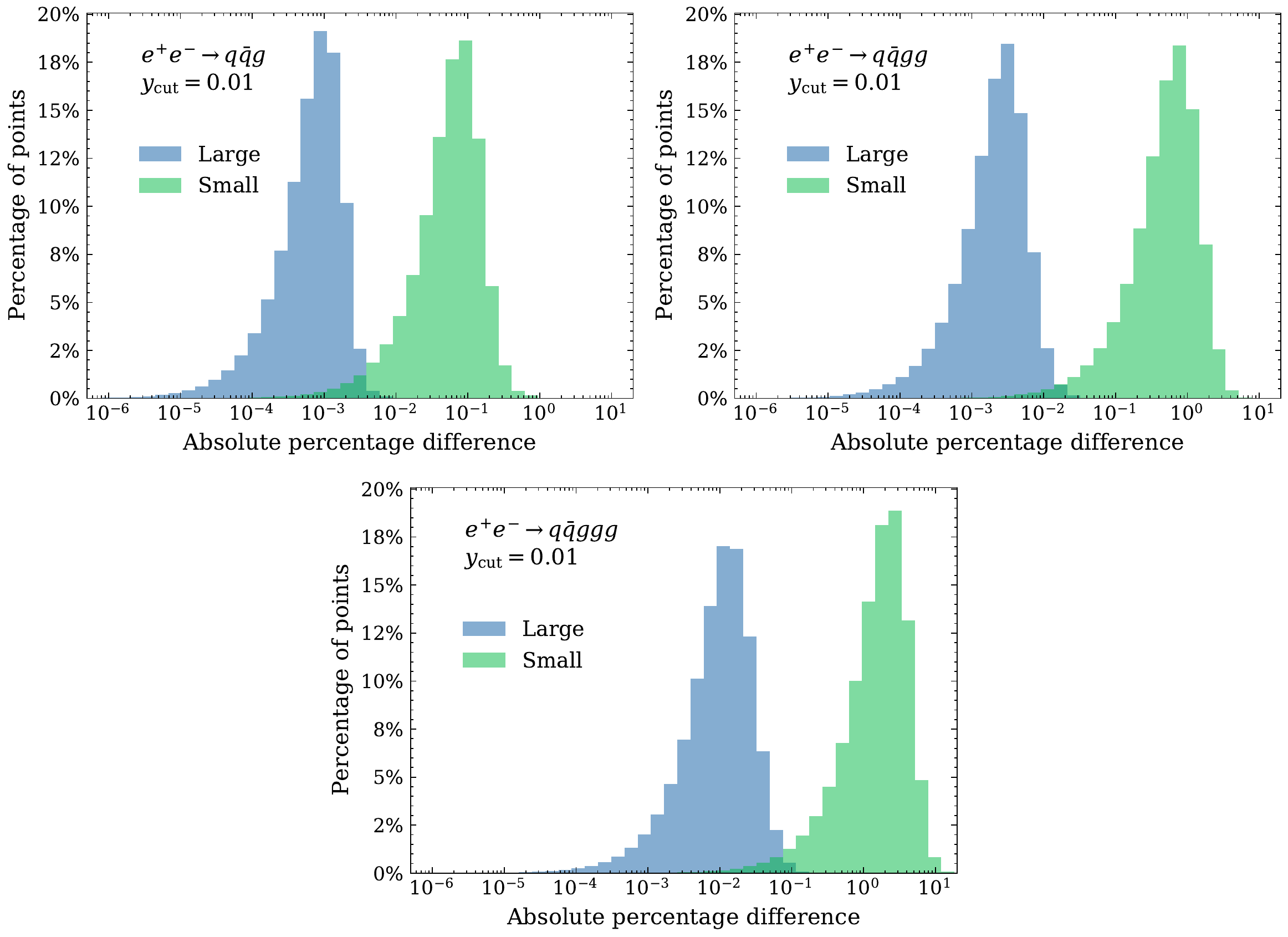}
    \caption{Comparison of error distributions of 10 million matrix element predictions between NNs used in Section~\ref{sec:n3jet_comparison} labelled as `Small' and our larger NNs labelled as `Large'.
    Note that `small' and `large' NNs were trained on 500k and 40m training samples, respectively.
    The testing data is identical to those shown in Figure~\ref{fig:main_error_distribution}, for the relevant global phase-space cuts.}
    \label{fig:error_500k_vs_40m}
\end{figure}

By increasing the size of the training datasets we aim to expose the neural network to more samples of the phase-space, thereby increasing accuracy on predictions on as much of the phase-space as possible.
Along with increasing the number of training samples, the neural network architecture has been expanded to include more hidden nodes and hidden layers.
The extra hidden nodes and layers introduces more parameters into the model allowing the neural network to utilise the additional data. 
We found that to get good performance, we had to balance the size of the training dataset used and the size of the network. i.e. a small network is not expected to capture all the variations in a large dataset as easily as a larger network would, due to the smaller number of parameters available to the network.
Of course, we also had to consider more physical constraints such as the time spent on training the neural networks which limits both the size of training datasets and architecture.
In Figure~\ref{fig:error_500k_vs_40m}, we show the improvements in accuracy of our main NNs compared to the smaller NNs from Section~\ref{sec:n3jet_comparison}.
Although the training and testing data for the smaller NNs are not identical due to differences in code used to generate the sets\footnote{although RAMBO is used for phase-space generation in both works, the selection criteria is different (JADE algorithm vs $k_{t}$ algorithm)}, for this comparison it suffices to show that the larger NNs are orders of magnitude more accurate than the smaller NNs.

\begin{figure}
    \centering
    \includegraphics[width=\textwidth]{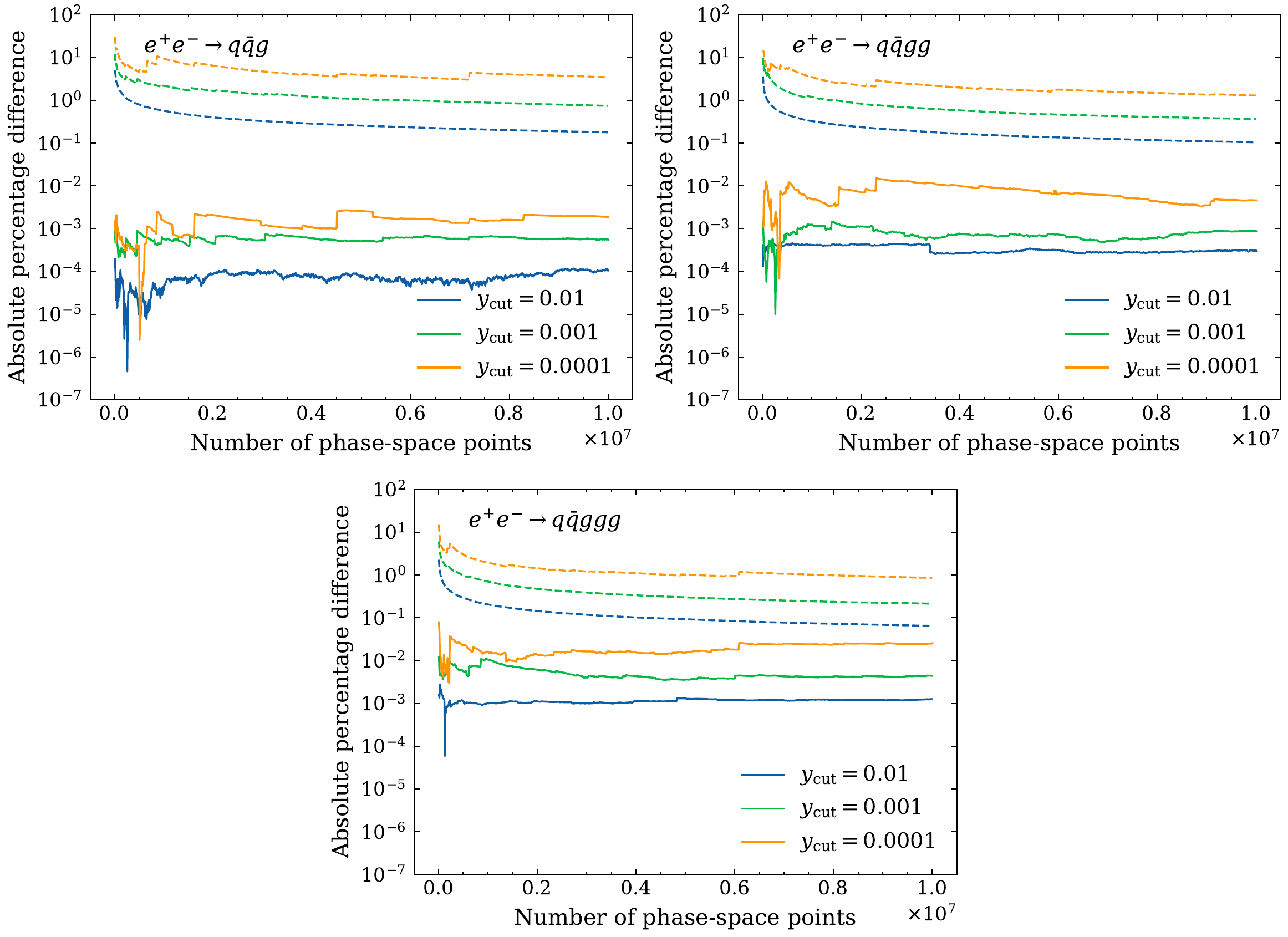}
    \caption{Absolute percentage differences between NN and {\NJet} cross-section predictions in solid lines.
    Dashed lines represent the Monte Carlo error expressed as a percentage error relative to the {\NJet} cross-section calculated with the corresponding number of phase-space points.
    Cross-sections have been calculated at intervals of 100k points up to the full 10 million phase-space points in the test set.
    Axis scales have been fixed to highlight the differences between multiplicities.}
    \label{fig:cross_section_diffs}
\end{figure}

Improvements in per-point accuracy translate to improved total cross-section predictions.
In Figure~\ref{fig:cross_section_diffs} we show the percentage differences of the NN cross-section predicted compared to those from {\NJet}.
There is a similar trend of errors increasing with increasing multiplicity and more inclusive phase-space cuts.
All total cross-section predictions are well below 0.1\% error.
There is a small systematic offset of the neural network cross-section compared to the {\NJet} cross-section that becomes apparent under closer inspection.
This was discussed in Ref.~\cite{Badger:2020uow} and we provide an additional explanation in Appendix~\ref{appendix:jensen}.

We also show in Figure~\ref{fig:cross_section_diffs} the estimated statistical Monte Carlo integration relative error for comparison.
The fact that the accumulated error on the matrix element is much lower than the MC error on the cross-section opens up the possibility 
to use the knowledge the network has gathered on the matrix element to augment the dataset to reduce the statistical error. Using such an 
augmentation technique would introduce a new systematic error on the prediction related to the accumulated network interpolation/extrapolation error, 
which would have to be balanced with the reduction in the MC integration error. Figure~\ref{fig:cross_section_diffs} suggests that the dataset could be 
augmented in such a way by a large factor before reaching a minimal overall uncertainty. This opportunity might not seem very useful for this particular example of leading 
order matrix elements where evaluations are relatively cheap computationally, but 
if a similar degree of accuracy in the emulation can be obtained for higher order matrix elements, this procedure could reduce the resource cost of matrix element calculation
significantly. We defer the study of this augmentation method to future work.

\begin{figure}
    \centering
    \includegraphics[width=\textwidth]{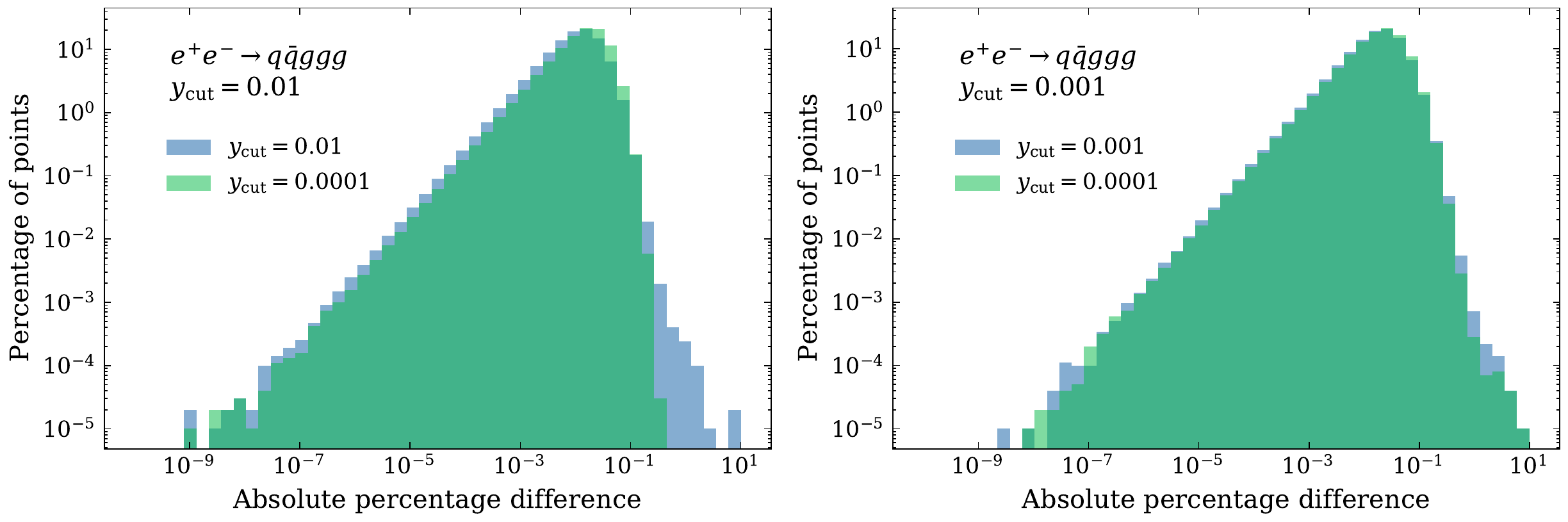}
    \caption{Left: error distribution on the $y_{\mathrm{cut}} = 0.01$ testing dataset as predicted by $y_{\mathrm{cut}} = 0.01$ and $y_{\mathrm{cut}} = 0.0001$ models.
    Right: error distribution on the $y_{\mathrm{cut}} = 0.001$ testing dataset as predicted by $y_{\mathrm{cut}} = 0.001$ and $y_{\mathrm{cut}} = 0.0001$ models.
    We use a logarithmic vertical axis to highlight the right-hand tail of the error distributions.}
    \label{fig:errors_relaxed_cuts}
\end{figure}
Since we retain good performance by relaxing the global phase-space cut, 
we carry out a simple test of generalisability by using the 5-jet $y_{\mathrm{cut}} = 0.0001$ model to infer on the two datasets with harsher cuts.
This is shown in Figure~\ref{fig:errors_relaxed_cuts} where we see that accuracy is comparable to the reference (blue) in both cases.
In the case of $y_{\mathrm{cut}} = 0.01$ (left), the model trained with more of the phase-space reduces errors in the right-hand tail of the distribution. 
This proves that enlarging the training phase-space can be done without having a large detrimental effect on the overall accuracy, and can significantly reduce the number of
large prediction errors.

\subsection{Random trajectories}\label{sec:trajectories}
\begin{figure}
    \centering
    \includegraphics[width=\textwidth]{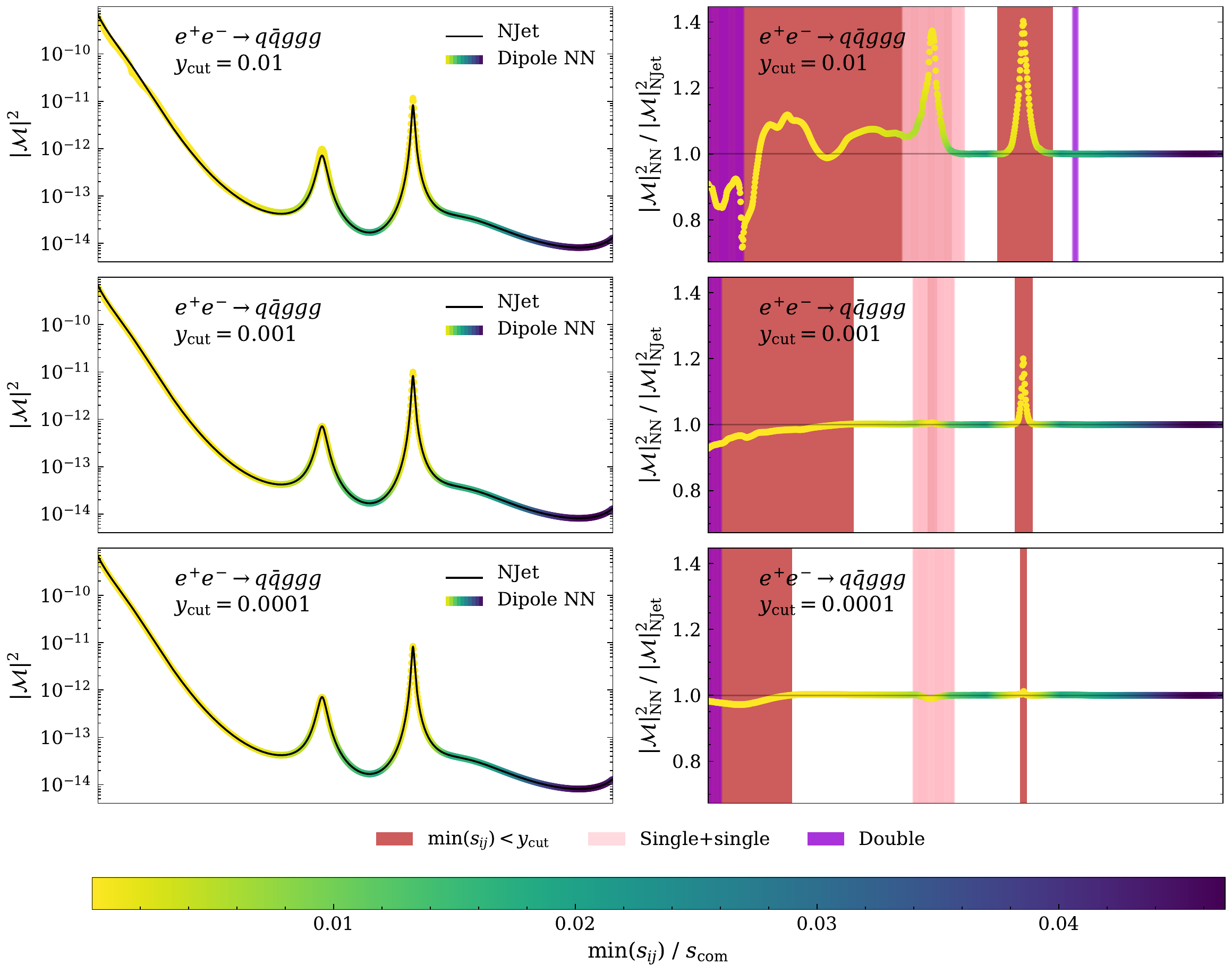}
    \caption{Left: matrix element prediction of random phase-space trajectory.
    Right: ratio of NN and {\NJet}.
    NN predictions are coloured by minimum $s_{ij}$ pair.
    Red bands indicate min($s_{ij}$) is smaller than $y_{\mathrm{cut}}$.
    Pink bands indicate where there are two separate single unresolved limits.
    Purple bands indicate double unresolved limits, i.e. three particles in one jet.
    Blue bands indicate where there is both a double and a single unresolved limit.
    The pink, purple, and blue bands represent regions of phase-space which would have been excluded by {\FastJet}.}
    \label{fig:random_trajectory}
\end{figure}

As another test of our emulator we assess the accuracy of our predictions on random phase-space trajectories.
These random phase-space trajectories are generated by connecting two random points in phase-space continuously without excluding any region of phase-space.
This presents an interesting and challenging test of the interpolation and extrapolation abilities of the NNs as some parts of the trajectories may lie outside of the phase-space region of the training datasets.
We show the results for 5-jet trajectories as this is the highest multiplicity we considered, predictions on lower multiplicities are better-behaved.
We investigated 50 different random trajectories\footnote{the corresponding plots are available in a document attached to this submission}. For the discussion in this article 
we chose one that contains many interesting features, namely the matrix elements span many orders of magnitudes and there are distinct peaks in the trajectory.
In Figure~\ref{fig:random_trajectory} we show the predictions by the three 5-jet models trained on data with different global phase-space cuts for this trajectory.
Left column shows the actual matrix element prediction and right column shows the ratios of the prediction to {\NJet}.
We analysed the predictions for the 50 random trajectories and measured the fraction of their length where the accuracy falls within given intervals.
Table~\ref{table:trajectory_predictions} shows the result for the regions of phase-space where training data was available, and those falling beyond the data available to the model. 

NN predictions are depicted as coloured scatter plots where the colour indicates the value of the minimum $s_{ij}$ between any pair of final-state particles at that phase-space point.
To more easily visualise the extrapolation performance of the NNs we highlight the regions where the minimum $s_{ij}$ goes below the global phase-space cut applied to the training set the models were trained on, 
for each cut made. The regions of the plots where this occurs have been coloured in red.  
With these trajectories being completely randomly selected in phase-space there is a possibility for there to be doubly singular points or worse.
To check for this we used {\FastJet} to cluster the phase-space points in the same way we did for data generation, see Section~\ref{sec:data}.
The pink regions indicate points which have two separate single unresolved limits, we label this configuration `Single+single'.
The purple regions indicate points which have a double unresolved limit (i.e. three particles in one jet), we label this configuration `Double'.
The blue regions indicate points which have both a double unresolved limit and a separate single unresolved limit, we label this configuration `Double+single'.
We do not include the quark-anti-quark invariant in defining these regions as there is no associated infrared divergence.
The pink, purple, and blue bands indicates regions of points which would have been discarded for our training and testing datasets.

Accuracy is high when the minimum $s_{ij}$ of the trajectory is not below any $y_{\mathrm{cut}}$, i.e. when the NN prediction curve is blue.
This is demonstrating that the NNs are interpolating well.
Performance generally declines in the red, pink, purple, and blue regions, which is not unexpected as the NNs are extrapolating.
Given that this trajectory has regions which go more collinear than any points the networks have been exposed to before, we would expect the networks which
have been trained with the smallest $y_{\mathrm{cut}}$ parameter to perform best.
We see that this is the case as accuracy is acceptable in the $y_{\mathrm{cut}} = 0.0001$ models, including in the regions where the NN is extrapolating.

\begin{table}
    \centering
    \begin{tabular}{|c|c|c|c|c|c|}
        \hline
        Region & $y_{\mathrm{cut}}$ & Fraction of pts & Pts outside $0.1\%$ & Pts outside $1\%$ & Pts outside $5\%$ \\
        \hline
        \multirow{3}{*}{White} & 0.01   & 36.5\% & 1.9\% & 0.0\% & 0.0\% \\
        \cline{2-6}
                               & 0.001  & 74.8\% & 0.98\% & 0.0\% & 0.0\% \\
        \cline{2-6}
                               & 0.0001 & 78.4\% & 1.9\% & 0.0\%  & 0.0\% \\
        \hline
        \hline
        \multirow{3}{*}{Pink}  & 0.01   & 27.6\% & 58.2\% & 25.2\% & 12.9\% \\
        \cline{2-6}
                               & 0.001  & 13.5\% & 31.7\% & 6.5\% & 0.53\% \\
        \cline{2-6}
                               & 0.0001 & 13.5\% & 38.9\% & 9.3\% & 2.8\% \\
        \hline
        \hline
        \multirow{3}{*}{Purple}& 0.01   & 7.2\% & 69.1\% & 32.1\% & 20.3\% \\
        \cline{2-6}
                               & 0.001  & 3.6\% & 31.3\% & 7.9\% & 4.5\% \\
        \cline{2-6}
                               & 0.0001 & 3.6\% & 30.5\% & 1.6\% & 0.0\% \\
        \hline
        \hline
        \multirow{3}{*}{Blue}  & 0.01   & 1.5\% & 79.3\% & 41.6\% & 0.77\% \\
        \cline{2-6}
                               & 0.001  & 0.4\% & 29.2\% & 0.0\% & 0.0\% \\
        \cline{2-6}
                               & 0.0001 & 0.4\% & 76.4\% & 0.0\% & 0.0\% \\
        \hline
        \hline
        \multirow{3}{*}{Red}   & 0.01   & 52.5\% & 76.2\% & 38.6\% & 20.5\% \\
        \cline{2-6}
                               & 0.001  & 7.0\%  & 69.9\% & 25.2\% & 9.5\% \\
        \cline{2-6}
                               & 0.0001 & 1.1\%  & 90.8\% & 33.8\% & 1.5\% \\
        \hline
    \end{tabular}
    \caption{The performance of trajectory predictions separated for white, pink, purple, blue, and red regions.
    We present the percentage of points that lie outside 0.1\%/1.0\%/5.0\% errors.
    Fraction of points indicates the percentage of points that lie in the region of interest, out of all phase-space points from the 50 trajectories we examine.
    }
    \label{table:trajectory_predictions}
\end{table}

In summary, we have shown that the neural networks show acceptable performance on random phase-space trajectories which are of different nature to the datasets used to train and test the networks.
Given that the general performance of the NNs of all three phase-space cuts are similar, it would make sense to use the models trained with the most inclusive phase-space cuts as it has been exposed
to more of the complete phase-space.

\section{Conclusion}\label{sec:conclusion}
In this article we presented a new strategy to emulate matrix elements using a neural network. By leveraging the knowledge of 
the factorisation properties of the matrix elements our model is able to extrapolate well outside of its training range. 
We showed that using this method we obtain significantly improved per-point accuracy than obtained in previous works. 
We also showed that the per-point accuracy of the model is not significantly affected by the generation cut for the training, which means that
it would be possible to train our emulator on very inclusive cuts, allowing them to be applied in a multitude of settings. 

The accuracy of the emulation could allow users to augment the training dataset to reduce the MC error of cross-sections or distributions while using 
fewer computing resources compared to the original calculation. We leave the investigation of this aspect to further work.

Our method was demonstrated in this article using a tree-level process, but it could be generalised to higher order matrix elements by adapting the set 
of ingredients made available to the network for the interpolation. Specifically, we can include extra terms into our ansatz to help capture additional divergences at higher orders. We look forward to applying this method to such cases.

\acknowledgments
We would like to thank Joseph Aylett-Bullock and Simon Badger for insightful comments on early drafts of the paper and for providing us with the data for our comparison.

\hspace{2cm}
\hrule

\appendix
\section{Azimuthal angle $\phi_{ij}$ calculation}\label{appendix:phi}
To calculate the azimuthal angle $\phi_{ij}$ for a pair of particle momenta $p_i$ 
and $p_j$ we first consider the plane perpendicular to the momentum 
\begin{eqnarray}\label{eq:pij}
    \vec{p}_{ij} = \vec{p}_{i} + \vec{p}_{j} \, .
\end{eqnarray}
We project the unit vector in the $z$ direction and the momentum of particle $i$ onto 
this plane\footnote{using particle $j$ instead results in a shift of $\phi_{ij}$ by 
$\pi$ which makes no difference for $\sin2\phi$ or $\cos2\phi$.}: 
\begin{eqnarray}\label{eq:north_east}
    \vec{r}_{z} &=& \vec{e}_{z} - \left(\dfrac{\vec{p}_{ij} \cdot \vec{e}_{z}}{\vec{p}_{ij}^{2}}\right) \vec{p}_{ij} \, , \\
    \vec{r}_{i} &=& \vec{p}_{i} - \left(\dfrac{\vec{p}_{ij} \cdot \vec{p}_{i}}{\vec{p}_{ij}^{2}}\right) \vec{p}_{ij} \,.
\end{eqnarray}
The angle $\phi_{ij}$ is the angle between these two projected vectors.
\begin{eqnarray}\label{eq:sin_cos}
    \sin{\phi_{ij}} = 
    \hat r_{ij}\cdot\left( \hat{r}_{i} \times \hat{r}_{z} \right)\, , \quad \cos{\phi_{ij}} = \hat{r}_{i} \cdot \hat{r}_{z} \, ,
\end{eqnarray}
where we have normalised all vectors to be unit vectors:
\begin{equation}
\hat r_z = \frac{\vec{r}_z}{|r_z|}\;, \qquad \hat r_i = \frac{\vec{r}_i}{|r_i|}\;, \qquad \hat r_{ij} = \frac{\vec{p}_{ij}}{|p_{ij}|}\;.
\end{equation}

\section{Jensen's Inequality}\label{appendix:jensen}
Jensen's inequality states that for concave functions
\begin{eqnarray}
    f(\mathbb{E}[Y]) \geq \mathbb{E}[f(Y)] \, ,
\end{eqnarray}
where $\mathbb{E}$ is the expectation value, the function $f$ in our case is $\arcsinh$ that behaves similarly to the natural logarithm due to the scale of the problem, and $Y$ is a random variable representing our target distribution.
This inequality can be rewritten as
\begin{eqnarray}
    f(\mathbb{E}[Y]) - \mathbb{E}[f(Y)] \geq 0 \, ,
\end{eqnarray}
which is known as Jensen's gap.
With a concave function, the mean of the transformed target distribution will always be underestimating the actual mean, i.e.
\begin{eqnarray}
    \mathbb{E}[Y] \geq \sinh{(\mathbb{E}[\arcsinh{(y)}])} \, ,
\end{eqnarray}
where the LHS is the actual expectation value (cross-section) of the random variable $Y$ and the RHS is the shifted expectation value that the neural network learns.
In our scenario the neural network reduces the variance of the residual distribution, meaning the gap in reality is small but there will always be an offset in the mean value learnt.

\section{Phase-space trajectories}\label{appendix:trajectory_generation}
We can see the RAMBO algorithm as a map from the unit hypercube in some high dimension into the $n$-particle phase-space. A flat distribution in the hypercube maps to a flat distribution in the multi-particle phase-space. 
To generate our phase-space trajectories we pick two points at random in the unit hypercube and map the line between them using the RAMBO mapping. As a result, each point on the resulting trajectory has equal probability density in phase-space. 
Any other phase-space generator that smoothly maps the unit hypercube to a multi-particle phase-space could replace RAMBO in this procedure and could lead to trajectories with very different characteristics. One could for example imagine a 
sophisticated algorithm that only maps points close to the boundary of the hypercube to soft or collinear configurations. Using such an algorithm would have trajectories that avoid configurations with many particles more collinear or soft than 
the end-points of the trajectories. We find that with RAMBO the trajectories tend not to avoid difficult phase-space configurations and are therefore a good test of the extrapolation properties of our method.

\begin{figure}
    \centering
    \includegraphics[width=0.45\textwidth]{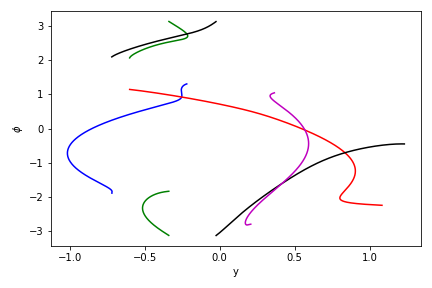}
    \includegraphics[width=0.45\textwidth]{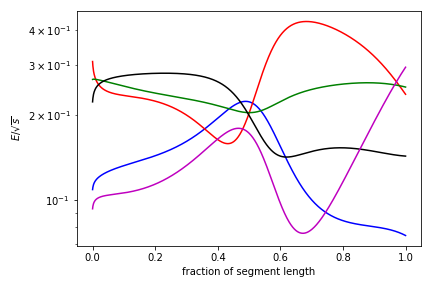}
    \caption{Left: rapidity and azimuthal angle trajectories for the five final state particles in the trajectories used in Section~\ref{sec:trajectories}.
    Right: Evolution of the same particle energies as a function of the position along the segment between the two random points in the unit hypercube.
    }
    \label{fig:detectorpath}
\end{figure}

Figure~\ref{fig:detectorpath} shows the trajectory we chose in Section~\ref{sec:trajectories}.

\bibliographystyle{JHEP}
\bibliography{nn_me}

\end{document}